\documentclass[pdflatex,sn-aps]{sn-jnl}

\usepackage{physics}
\usepackage{amsmath,amssymb,amsfonts}
\usepackage[english]{babel}
\usepackage[utf8]{inputenc}
\usepackage[T1]{fontenc}
\usepackage{mathrsfs}
\usepackage{bbm}
\usepackage{bm}
\usepackage{enumerate}
\usepackage{graphicx}
\usepackage[dvipsnames]{xcolor}
\usepackage{float}
\usepackage{subdepth}
\usepackage{fnpct}
\usepackage{booktabs}
\usepackage{multirow}
\usepackage[math]{cellspace}
\usepackage[title]{appendix}%
\usepackage{manyfoot}
\usepackage{anyfontsize}
\usepackage{ragged2e}
\usepackage{lipsum}
\usepackage{comment}

\newcommand{\av}[1]{\langle#1\rangle}

\renewcommand{\(}{\left(}
\renewcommand{\)}{\right)}
\newcommand{\id}{\mathbbm{1}}

\newcommand{\scT}{\mathcal{T}}

\newcommand{\At}{\widetilde{A}}
\newcommand{\Bt}{\widetilde{B}}
\newcommand{\Ct}{\widetilde{C}}

\newcommand{\Jt}{\widetilde{J}}
\newcommand{\phit}{\tilde{\phi}}
\newcommand{\dsP}{\mathbb{P}}
\newcommand{\at}{\widetilde{a}}

\newcounter{ls}

\newcounter{pr}

\begin{document}
	
\title{Symmetry classes of classical stochastic processes}

\author*[1]{Lucas  S\'a}
\email{lucas.seara.sa@tecnico.ulisboa.pt}

\author[2,3]{Pedro Ribeiro}
\email{ribeiro.pedro@tecnico.ulisboa.pt}

\author[4,5]{Toma\v z Prosen}
\email{tomaz.prosen@fmf.uni-lj.si}

\author[6]{Denis Bernard}
\email{denis.bernard@ens.fr\vspace{-2ex}}

\affil*[1]{TCM Group, Cavendish Laboratory, University of Cambridge, JJ Thomson Avenue, Cambridge CB3 0HE, UK}
\affil[2]{CeFEMA, Instituto Superior T\'ecnico, Universidade de Lisboa, Av.\ Rovisco Pais, 1049-001 Lisboa, Portugal}
\affil[3]{Beijing Computational Science Research Center, Beijing 100193, China}
\affil[4]{Department of Physics, Faculty of Mathematics and Physics, University of Ljubljana, SI-1000 Ljubljana, Slovenia}
\affil[5]{Institute of Mathematics, Physics and Mechanics, SI-1000 Ljubljana, Slovenia}
\affil[6]{Laboratoire de Physique de l'\'Ecole Normale Sup\'erieure, CNRS, ENS \& PSL University, Sorbonne Universit\'e, Universit\'e Paris Cit\'e, 75005 Paris, France\vspace{-3ex}}

\abstract{
    We perform a systematic symmetry classification of the Markov generators of classical stochastic processes. Our classification scheme is based on the action of involutive symmetry transformations of a real Markov generator, extending the Bernard-LeClair scheme to the arena of classical stochastic processes and leading to a set of up to fifteen allowed symmetry classes. We construct families of solutions of arbitrary matrix dimensions for five of these classes with a simple physical interpretation of particles hopping on multipartite graphs. In the remaining classes, such a simple construction is prevented by the positivity of entries of the generator particular to classical stochastic processes, which imposes a further requirement beyond the usual symmetry classification constraints. We partially overcome this difficulty by resorting to a stochastic optimization algorithm, finding specific examples of generators of small matrix dimensions in six further classes, leaving the existence of the final four allowed classes an open problem. Our symmetry-based results unveil new possibilities in the dynamics of classical stochastic processes: Kramers degeneracy of eigenvalue pairs, dihedral symmetry of the spectra of Markov generators, and time reversal properties of stochastic trajectories and correlation functions.
}
 
\maketitle

{
\hypersetup{linkcolor=black}
\tableofcontents
\markboth{\leftmark}{\rightmark}
}

\clearpage

\section{Introduction}

Random matrix theory pervades modern physics and mathematics~\cite{akemann2011oxford}. One of its most appealing features is that spectral correlations of complex systems depend only on a small number of symmetries and coincide with those of a random matrix with the same symmetries. Consequently, organizing systems into classes labeled by their symmetries provides extensive information from minimal input. Famous examples include the degree of repulsion between energy levels, the double degeneracy of half-integer-spin time-reversal-symmetric systems, or the universality of conductance fluctuations. These are often the only analytical statements one can make for systems that are not solvable.
Quantum mechanics still provides not only most of the applications of random matrix theory but also the most common framework to classify random matrices. Indeed, work on chiral fermions in QCD~\cite{verbaarschot1993PRL,verbaarschot1994PRL} and disordered superconductors~\cite{altland1997} established a tenfold classification of Hermitian Hamiltonians (or, more generally, random matrices) based on unitary and antiunitary symmetries (time-reversal, particle-hole, and chiral), the Altland-Zirnbauer classes. More recently, symmetry classifications of non-Hermitian quantum dynamical generators have also been put forward, for the cases of general non-Hermitian Hamiltonians~\cite{bernard2002,kawabata2019PRX,garcia2022PRX}, PT-symmetric Hamiltonians~\cite{garcia2023ARXIVb}, and Lindbladians~\cite{sa2023PRX,kawabata2023PRXQ,altland2021PRX,lieu2020PRL,kawasaki2022PRB}.

Despite the power and generality of the symmetry classification approach, it has remained mostly circumscribed to the quantum realm. A natural question is, then, how to extend it to classical settings---in particular, nonequilibrium ones described by non-Hermitian matrices---, raising the possibility of previously unidentified classes of classical stochastic dynamics, with unique spectral and dynamical consequences. We will focus our attention on classical Markov processes, which describe systems exhibiting memoryless and probabilistic transitions between states, with applications ranging from physical, chemical, and biological systems, to economics.
The solution of a classical Markov process is obtained by specifying the time-evolution of a (real) probability vector $\bm{p}$ ($\sum_{i=1}^N p_i=1$) of an $N$-dimensional system, governed by the Markov equation
\begin{equation}
\partial_t \bm{p}=L \bm{p}.
\end{equation}
Conservation of probability requires that the Markov generator, an $N\times N$ matrix $L$, satisfies the three properties:
\begin{align}
	\label{eq:def_Markov_1}
	&L_{ij}\in \mathbb{R},\\
	\label{eq:def_Markov_2}
	&L_{ij}\geq 0, \quad i\neq j,\\
	\label{eq:def_Markov_3}
	&\sum_{i=1}^N L_{ij}=0, \quad \forall j.
\end{align}
Such processes are a natural candidate to investigate symmetry classifications in classical systems, as they can be obtained as a limit (diagonal projection) of quantum Lindbladian evolution, for which the symmetry classification is well-established~\cite{sa2023PRX}.

In this paper, we show that indeed symmetry classifications are not restricted to the quantum realm, and can be extended to classical stochastic processes. In particular, we show that considering (anti)unitary symmetries is not crucial to developing the classification, thus extending the Altland-Zirnbauer~\cite{altland1997} and Bernard-LeClair schemes~\cite{bernard2002,kawabata2019PRX} to general involutive symmetries. We then apply our classificatory scheme to the Markov generators of classical stochastic processes, finding that they can belong at most to fifteen different symmetry classes (Sec.~\ref{sec:classification}). Interestingly, a certain constraint of the quantum symmetry classification that prevents classes with Kramers degeneracy (the exact double degeneracy of every eigenvalue)~\cite{sa2023PRX} is absent in the classical limit, illustrating a rich interplay of symmetry and classical stochastic evolution going beyond a mere projecting of quantum Markov dynamics to the classical subspace. 
Notwithstanding, the classical setting presents its own set of difficulties for the classification. Indeed, contrary to the quantum case, the existence of a preferred basis (where probabilities are real and positive) is at odds with the basis-independent statements sought after by symmetry classifications. Specifically, while we can explicitly build parametric families of Markov generators belonging to five of the fifteen classes, the aforementioned tension prevents the construction of examples in the remaining ten classes if working in the computational basis (Sec.~\ref{sec:families}). To overcome this difficulty, we developed a stochastic optimization algorithm, which allows us to find examples in six further classes (Sec.~\ref{sec:systematic}). The status of the other four classes remains inconclusive, as we could neither build an example for any of them nor exclude its existence from our general fifteenfold classification. Finally, we also discuss consequences for the spectrum of the Markov matrix and for the dynamics, such as Kramers degeneracy of pairs of eigenvalues, dihedral symmetry of the spectrum, time reversal of trajectories and correlation functions, and the role of detailed balance (Sec.~\ref{sec:physical_consequences}).

\section{Fifteenfold symmetry classification}
\label{sec:classification}

In quantum mechanics, to perform the symmetry classification of a non-Hermitian matrix, we look for the antiunitary and unitary symmetry involutions of its irreducible blocks. In a classical stochastic process, there is, \textit{a priori}, no reason to constrain the symmetries to be unitary or antiunitary. Moreover, the Markov generator $L$ is automatically real and, without any reason to introduce complex numbers at this stage, we therefore demand that the operators implementing the symmetries are themselves real. In what follows, we extend the Bernard-LeClair classification of non-Hermitian matrices~\cite{bernard2002,kawabata2019PRX} to this setting.

If $L$ has a commuting symmetry,
\begin{equation}
	U L U^{-1} = L,
\end{equation} 
then, $L$ and $U$ can be simultaneously diagonalized. Inside each block of $L$, $U$ acts trivially as a multiple of the identity. If no further commuting symmetries exist, we call such a block an irreducible block. The classification of $L$ has to be done for each irreducible block individually.
Importantly, the rotation into the diagonal basis of the commuting symmetries $U$ does not in general preserve the Markov properties~(\ref{eq:def_Markov_1})--(\ref{eq:def_Markov_3}) in each of the irreducible blocks. For example, most blocks are not represented by a real matrix if $U$ is not itself real with real eigenvalues. Therefore, the classification we develop below is strictly valid only in the cases where no commuting unitary symmetries exist. It is, however, natural to conjecture that the block containing the invariant measure of the Markov process (henceforth referred to as the steady-state block) can still be recast into Markov form. Indeed, consider an initial probability vector fully supported in the steady-state block. Under the Markov dynamics, the initial probability vector evolves into another admissible probability vector. But because all blocks of $L$ are decoupled, no component of the probability vector outside of the steady-state block is generated. We conclude that the steady-state block maps probability vectors into probability vectors and it should be possible to write it in Markov form. 
In App.~\ref{app:Markov_steady_state} we prove this statement under some mild assumptions about the projection into the steady-state block and give an illustrative example of a Markov matrix with SU(2) symmetry.
We thus proceed with the classification assuming that either $L$ has no commuting symmetries or, if it has, that we restrict attention to the steady-state block.

We define the (traceless) shifted Markov generator
\begin{eqnarray}
	L'=L-\frac{\Tr L}{\Tr \id}\id,
\end{eqnarray}
(it is clear that a commuting symmetry $U$ of $L$ is also one of $L'$ and their irreducible blocks coincide),
and look for the possible existence of the following symmetries:
\begin{align}
	&S L' S^{-1} = - L', \label{eq:def_S}\\
	&R_+ L'^T R_+^{-1} = L', \label{eq:def_Rp}\\
	&R_- L'^T R_-^{-1} =  - L',  \label{eq:def_Rm}
\end{align}
where $S$ and $R_\pm$ are invertible real matrices only constrained by 
\begin{align}
	&S^2=\eta_S, \label{eq:square_S}\\
	&R_+ R_+^{-T}=\eta_+, \label{eq:square_Rp}\\
	&R_- R_-^{-T}=\eta_-, \label{eq:square_Rm}\\
	&R_+ R_-^{-T}=\epsilon R_- R_+^{-T}, \label{eq:comm_RpRm}
\end{align}
where $\eta_S,\eta_\pm,\epsilon\in\{\pm1\}$. Here, $()^T$ denotes transposition, and $R^{-T}$ is a shorthand notation for $R^{-T}:=(R^T)^{-1}=(R^{-1})^T$. We do not consider the existence of more than one symmetry of each type, since their composition commutes with $L'$ and we assumed $L'$ to be irreducible. Moreover, the three transformations are not independent, as the composition of any two yields the third. Consequently, we must consider the presence of either only one of them or all three. In the latter case, we take $R_\pm$ as the independent symmetries and $S=R_+ R_-^{-T}$ as the derived operation.

The constraints~(\ref{eq:square_S})--(\ref{eq:comm_RpRm}) generalize the restrictions imposed on the symmetry operators in quantum mechanics, where $R_\pm$ and $S$ must be unitary and $S$ always squares to $+1$. On the other hand, we have imposed the reality of $L'$, $R_\pm$, and $S$, which is absent otherwise.
The constraints are derived from the irreducibility of $L'$, by successive application of multiple symmetries, as we now show. To this end, we note that Eqs.~(\ref{eq:def_S})--(\ref{eq:def_Rm}) are invariant under $X\mapsto \gamma X$, where $X=S,R_\pm$ and $\gamma$ is a real number, and, therefore, we can freely rescale $X$ (we emphasize there is no phase freedom). We start by applying the $S$ symmetry twice: $S^2 L' S^2=L'$. By irreducibility of $L'$, $S^2$ must be a multiple of the identity, $S^2=\lambda \id$, with $\lambda$ real, and redefining $S\mapsto S/\sqrt{|\lambda|}$, we arrive at Eq~(\ref{eq:square_S}). Similarly, acting twice with, say, $R_+$ yields $R_+ R_+^{-T}=\lambda \id$ for some real $\lambda$. It follows that $R_+ = \lambda R_+^T=\lambda(\lambda R_+^T)^T=\lambda^2 R_+$ and, hence, $\lambda=\pm1$, implying Eq.~(\ref{eq:square_Rp}). The same applies to $R_-$. Normalizing the ``squares'' of $R_\pm$ did not require using their scale freedom, which can be used to fix the ``commutation relation'' of $R_+$ and $R_-$ instead. Indeed, applying them successively to $L'$, we have $R_+(R_- L'^T R_-^{-1})^T R_+^{-1}=-L'=R_-(R_+ L'^T R_+^{-1})^T R_-^{-1}$. The irreducibility of $L'$ then implies that $R_+ R_-^{-T}=\lambda R_- R_+^{-T}$ for a real $\lambda$. Redefining $R_+\mapsto \sqrt{|\lambda|}R_+$, we arrive at Eq.~(\ref{eq:comm_RpRm}). This exhausts the scale freedom of the symmetry operations. Finally, we recall that, in the presence of both $R_\pm$, $S=R_+ R_-^{-T}$ is a derived operation. In particular, its square is determined by the ``squares'' of $R_\pm$ and their ``commutation relation'', namely, 
\begin{equation}
    \eta_S=\epsilon\eta_+\eta_-,
\end{equation}
and we can thus use them interchangeably.

\begin{table}[t]
    \caption{\justifying Fifteenfold classification of classical Markov matrices. The first column lists the name of the class. (We adapted the nomenclature of Ref.~\cite{kawabata2019PRX} to the present setting; note that while we follow their conventions, our notations differ slightly.) The following three columns give the values of $\eta_+$, $\eta_-$, and $\eta_S$, respectively. (The two indices of the Cartan labels in the class names are, respectively, $\eta_S$ and $\eta_+\eta_-$.) The column ``Analytical'' indicates whether we have an analytic parametrization of a family of Markov matrices in the respective class (Sec.~\ref{sec:families}), while the column ``Numerical'' indicates whether we have found numerical examples of generators in that class using our Monte Carlo sampling procedure (Sec.~\ref{sec:systematic}). For the latter, in parentheses, we indicate whether the cost function of the numerical solution reaches zero in a finite number of steps (``$f=0$'') or goes down toward zero with the number of iterations (``$f\to0$''). The last two columns specify whether the spectra of generators in the respective class have some special properties (Sec.~\ref{sec:physical_consequences}), namely, dihedral symmetry (column ``Dihedral'') and/or Kramers degeneracy (column ``Kramers'').}
	\begin{tabular}{@{}lcccllll@{}}
		\toprule
		Class          & $R_+R_+^{-T}$ & $R_-R_-^{-T}$ & $S^2$ & Analytical & Numerical     & Dihedral & Kramers \\ 
            \midrule
		AI             & $0$           & $0$           & $0$   & Yes        & Yes ($f=0$)   & No       & No      \\
		AI$_+$         & $0$           & $0$           & $+1$  & Yes        & Yes ($f=0$)   & Yes      & No      \\
		AI$_-$         & $0$           & $0$           & $-1$  & No         & Yes ($f=0$)   & Yes      & No      \\
		BDI$^\dagger$  & $+1$          & $0$           & $0$   & Yes        & Yes ($f=0$)   & No       & No      \\
		DIII$^\dagger$ & $-1$          & $0$           & $0$   & No         & Yes ($f\to0$) & No       & Yes     \\
		BDI            & $0$           & $+1$          & $0$   & No         & No            & Yes      & No      \\
		CI             & $0$           & $-1$          & $0$   & No         & Yes ($f=0$)   & Yes      & No      \\
		BDI$_{++}$     & $+1$          & $+1$          & $+1$  & Yes        & No            & No       & No      \\
		CI$_{+-}$      & $+1$          & $-1$          & $+1$  & Yes        & Yes ($f=0$)   & No       & No      \\
		BDI$_{+-}$     & $-1$          & $+1$          & $+1$  & No         & No            & No       & Yes     \\
		CI$_{++}$      & $-1$          & $-1$          & $+1$  & No         & Yes ($f\to0$)   & No       & Yes     \\
		BDI$_{-+}$     & $+1$          & $+1$          & $-1$  & No         & No            & No       & No      \\
		CI$_{--}$      & $+1$          & $-1$          & $-1$  & No         & Yes ($f=0$) & No       & No      \\
		BDI$_{--}$     & $-1$          & $+1$          & $-1$  & No         & No            & No       & Yes     \\
		CI$_{-+}$      & $-1$          & $-1$          & $-1$  & No         & Yes ($f\to0$) & No       & Yes     \\
            \bottomrule
	\end{tabular}
	\label{tab:classes}
\end{table}

The symmetry class of (an irreducible block of) $L$ is thus determined by three signed numbers $\pm 1$, the ``squares'' of the symmetry operations, $S^2$ and $R_\pm R_\pm^{-T}$. Counting them is now a simple matter (they are summarized in Table~\ref{tab:classes}):
\begin{itemize}
	\item \textit{No symmetry.} One class, dubbed AI.
	\item \textit{One symmetry.} We have three types of symmetries, each with two possible values for the ``square'', leading to $3\times 2=6$ classes, dubbed AI$_+$, AI$_-$, BDI$^\dagger$, DIII$^\dagger$, BDI, and CI.
	\item \textit{Three symmetries.} Each of the three symmetries $R_\pm$ and $S$ can ``square'' to two values, giving $2^3=8$ classes, dubbed BDI$_{++}$, BDI$_{+-}$, BDI$_{-+}$, BDI$_{--}$, CI$_{++}$, CI$_{+-}$, CI$_{-+}$, and CI$_{--}$ (alternatively, as mentioned above, we could specify whether $R_+$ and $R_-$ ``commute'' or ``anticommute'' instead of the ``square'' of $S$).
\end{itemize}
In total, we thus have $1+6+8=15$ possible symmetry classes in classical stochastic processes. They are in one-to-one correspondence with the 15 non-Hermitian Bernard-LeClair classes with $\scT_+^2=-1$ (from which we borrowed their names)~\cite{kawabata2019PRX}, but are implemented by nonunitary and nonantiunitary real matrices that satisfy the generalized constraints~(\ref{eq:square_S})--(\ref{eq:comm_RpRm}). 

The allowed set of classes followed simply by restricting our attention to real matrices with real symmetry operations [Eq.~(\ref{eq:def_Markov_1})] and has not yet taken into account the restrictions of Eqs.~(\ref{eq:def_Markov_2}) and (\ref{eq:def_Markov_3}). In particular, it is not guaranteed that all 15 classes can be realized with those restrictions. In the next two sections, we look for explicit examples of Markov matrices in each of these classes. First, in Sec.~\ref{sec:families} we will follow the usual procedure~\cite{haake2013,magnea2008,sa2023PRX} and look for analytic parameterizations of families of matrices of arbitrary dimension. However, because of the Eqs.~(\ref{eq:def_Markov_2}) and (\ref{eq:def_Markov_3}), we can only do so in five of the fifteen classes. Therefore, we then turn to obtain further examples by Monte Carlo sampling in Sec.~\ref{sec:systematic}, which yields examples in ten classes. Remarkably, there is one class with an analytic expression that is \emph{not} found by our numerical sampling algorithm. Therefore, we establish that the symmetry classification of Markov generators subjected to Eqs.~(\ref{eq:def_Markov_2}) and (\ref{eq:def_Markov_3}) is \emph{at least} elevenfold but because of this limitation of the Monte Carlo algorithm, we cannot definitively rule out the existence of the remaining four.

\section{Analytical families of Markov matrices}
\label{sec:families}

We first give explicit realizations of Markov matrices for five classes, namely, AI, AI$_+$, BDI$^\dagger$, BDI$_{++}$, and CI$_{+-}$. These simple descriptions are possible because in those cases there is no mismatch between the preferred computational basis, in which the positivity constraints are simple, and the basis in which the operators implementing the discrete symmetries are simple. Borrowing the notation to be introduced in Sec.~\ref{sec:systematic}, these cases correspond to those for which $W=\id$.

To this end, we parametrize the classical Markov generator $L$ in terms of a real matrix $M$ with arbitrary diagonal entries and nonnegative off-diagonal entries:
\begin{equation}
	\label{eq:Markov_parametrization}
	L_{ij}=M_{ij}-\delta_{ij} \sum_{k=1}^N M_{kj},
	\qquad
	M_{ij}\geq 0, \quad i\neq j.
\end{equation}
Without loss of generality, we take $M$ to be traceless. To ensure that $L$ has the appropriate symmetry, we demand that $M$ satisfies the $S$ or $R_\pm$ symmetries, accordingly:
\begin{align}
\label{eq:symm_T-}
&S M S^{-1}=-M,
\\
\label{eq:symm_C-}
&R_- M^T R_-^{-1}=-M,
\\
\label{eq:symm_C+}
&R_+ M^T R_+^{-1}=+M.
\end{align}
However, this is not sufficient to guarantee a symmetry of $L$. 
We further impose the following necessary condition:
\begin{equation}
\label{eq:symm_cond_T-}
\sum_{n}M_{nk}=-\frac{\alpha}{2}, \quad \forall k,
\end{equation}
with $\alpha=\frac{2}{N}\Tr L$. Equation~\eqref{eq:symm_cond_T-} is a very restrictive condition as it demands that summing matrix elements of $M$ on each column yields the same output. Notwithstanding, it amounts only to a gauge choice ensuring that a symmetry of $M$ is passed to $L$ and does not restrict the physical processes. To see this, notice that we can shift $M$ by any diagonal matrix, i.e., $M_{ij}\to \widetilde M_{ij}= M_{ij} - \delta_{ij} x_j$ gives the same matrix $L$. The gauge degrees of freedom $x_j$ are exactly enough to always impose that both $\Tr M=0$ and $\sum_k M_{kj}=-\alpha/2$. Finally, note that, assuming Eq.~\eqref{eq:symm_cond_T-}, the constant $\alpha$ is determined by summing over $k$: since $\Tr M=0$ by gauge choice, Eq.~\eqref{eq:Markov_parametrization} implies $\Tr L= -\sum_{kj} M_{kj}$.

Let us first consider the case of an $S$ symmetry. We have:
\begin{equation}
\label{eq:symm_A_T-}
\begin{split}
\sum_{kl} S_{ik} L_{kl} (S^{-1})_{lj}
&=\sum_{kl} S_{ik} M_{kl} (S^{-1})_{lj} 
-\sum_{kl} S_{ik} \delta_{kl}\sum_{n}M_{nl} (S^{-1})_{lj} 
\\
&=-M_{ij}-\sum_{k}S_{ik}(S^{-1})_{kj}\sum_{n}M_{nk}. 
\end{split}
\end{equation}
To have an $S$ symmetry, we need, in addition to Eq.~(\ref{eq:symm_T-}), that
\begin{equation}
\sum_{k}S_{ik}(S^{-1})_{kj}\sum_{n}M_{nk}=-\delta_{ij}\sum_{n}M_{nj},
\end{equation}
which is achieved if Eq.~(\ref{eq:symm_cond_T-}) holds.
Then, Eq.~(\ref{eq:symm_A_T-}) reads as:
\begin{equation}
\label{eq:sym_A_T-_2}
\sum_{kl} S_{ik} L_{kl} (S^{-1})_{lj}
=-M_{ij}+\frac{\alpha}{2}\delta_{ij}
=-M_{ij}+\delta_{ij}\sum_{n}M_{nj}+\alpha \delta_{ij}
=-L_{ij}+\alpha \delta_{ij},
\end{equation}
which is Eq.~(\ref{eq:def_S}) for the shifted Markov matrix.

Next, we consider the case of a $R_\pm$ symmetry. As before, we have:
\begin{equation}
\label{eq:symm_A_C-}
    \sum_{kl} (R_\pm)_{ik} L_{kl} (R_\pm^{-1})_{lj}
    =\pm M_{ij}-\sum_{k}(R_\pm)_{ik}(R_\pm^{-1})_{kj}\sum_{n}M_{nk}. 
\end{equation}
We notice that the same condition~(\ref{eq:symm_cond_T-}) is sufficient to guarantee the $R_\pm$ symmetry: for $R_-$, the reasoning is exactly the same as above; for $R_+$, we note that
\begin{equation}
\sum_{kl} (R_+)_{ik} L_{kl} (R_+^{-1})_{lj}
=M_{ij}+\frac{\alpha}{2}\delta_{ij}
=M_{ij}-\delta_{ij}\sum_{n}M_{nj}
=L_{ij}.
\end{equation}

We can now follow the procedure of Ref.~\cite{magnea2008} and seek the simplest irreducible parametrization of the elements of each symmetry class that is consistent with the symmetries of that class, Eqs.~(\ref{eq:symm_T-})--(\ref{eq:symm_C+}), and the positivity of the entries. By irreducible we again mean that no additional unitary symmetries exist. Equation~(\ref{eq:symm_cond_T-}) has to be imposed on top of the parametrizations we derive.
Throughout, $X,Y,Z$ denote the three real traceless $2\times 2$ matrices
\begin{equation}
X=\begin{pmatrix}
		0 & 1 \\ 1 & 0
	\end{pmatrix},\quad
 Y=\begin{pmatrix}
		0 & 1 \\ -1 & 0
	\end{pmatrix},\quad
 Z=\begin{pmatrix}
		1 & 0 \\ 0 & -1
	\end{pmatrix},
\end{equation} 
$\id$ is the $N$-dimensional identity and $A$, $B$, $C,\dots$ are $N\times N$ real matrices (arbitrary, unless specified otherwise).

\subsection{Class AI, $S^2=R_+R_+^{-T}=R_-R_-^{-T}=0$}
\label{subsec:class_AI}

There are no symmetries to implement. $M$ is an arbitrary real matrix with nonnegative off-diagonal entries.

\subsection{Class AI$_+$, $S^2=+1$, $R_+R_+^{-T}=R_-R_-^{-T}=0$}
\label{subsec:class_AIp}

Class AI$_{+}$ is labeled by the symmetry $S^2=+1$. The simplest nontrivial choice of $S$ satisfying $S^2=+1$ and compatible with the reality of the Markov matrix is $S=Z\otimes \id$. In this case, $M$ is an $2N\times 2N$ matrix with block structure
\begin{equation}
	\label{eq:M_block}
	M=\begin{pmatrix}
		A & B \\ C & D
	\end{pmatrix}.
\end{equation}
Equation~(\ref{eq:symm_T-}) reads as
\begin{equation}
	S M S^{-1}
	=
	\begin{pmatrix}
		\id & 0 \\ 0 & -\id
	\end{pmatrix}
	\begin{pmatrix}
		A & B \\ C & D
	\end{pmatrix}
	\begin{pmatrix}
		\id & 0\\ 0 & -\id
	\end{pmatrix}
	=
	\begin{pmatrix}
		A & -B\\ -C & D
	\end{pmatrix}
	=
	\begin{pmatrix}
		-A & -B\\ -C & -D
	\end{pmatrix}
	=-M,
\end{equation}
from which it follows that the diagonal blocks vanish ($A=D=0$) because of the positivity of the entries, while there are no further conditions on the off-diagonal blocks (and we can rename $B$ to $A$ and $C$ to $B$). 

To apply Eq.~(\ref{eq:symm_cond_T-}), we define
\begin{equation}
	\label{eq:tilde_transf}
	\Jt_{ij}=\frac{J_{ij}}{\sum_{k}J_{kj}}
\end{equation}
for an arbitrary matrix $J$. The parametrization of a Markov generator in class AI$_+$ is finally found to be:
\begin{equation}
	\label{eq:parametrization_AI+}
	M=\begin{pmatrix}
		0 & \At \\ \Bt & 0
	\end{pmatrix}.
\end{equation}

\subsection{Class BDI$^\dagger$, $R_+R_+^{-T}=+1$, $R_-R_-^{-T}=S^2=0$}
\label{subsec:class_BDId}

Class BDI$^\dagger$ is labeled by the symmetry $R_+R_+^{-T}=+1$. Taking $R_+=X\otimes\id$ and $M$ of the form~(\ref{eq:M_block}), Eq.~(\ref{eq:symm_C+}) reads as
\begin{equation}
	R_+ M^T R_+^{-1}
	=
	\begin{pmatrix}
		0 & \id \\ \id & 0
	\end{pmatrix}
	\begin{pmatrix}
		A^T & C^T \\ B^T & D^T
	\end{pmatrix}
	\begin{pmatrix}
		0 & \id \\ \id & 0
	\end{pmatrix}
	=
	\begin{pmatrix}
		D^T & B^T\\ C^T & A^T
	\end{pmatrix}
	=
	\begin{pmatrix}
		A & B\\ C & D
	\end{pmatrix}
	=M,
\end{equation}
whence it follows that
\begin{equation}
	\label{eq:parametrization_BDIdg}
	M=\begin{pmatrix}
		A & B \\ C & A^T
	\end{pmatrix},
	\qquad
	\begin{cases}
		B=B^T\\C=C^T
	\end{cases}.
\end{equation}

Now, to apply Eq.~(\ref{eq:symm_cond_T-}), we cannot simply renormalize the matrix elements by replacing $A$, $B$, and $C$ by $\At$, $\Bt$, and $\Ct$ because the transformation~(\ref{eq:tilde_transf}) spoils the transposition symmetry ($\Bt\neq \Bt^T$, etc.). Instead, we can proceed as follows. We can choose the entries of the upper-left block $A$ freely. This also fixes the lower-right block $A^T$. We can also choose arbitrary upper-triangular entries for $B$ and $C$; $B=B^T$ and $C=C^T$ then fixes the lower-triangular entries. Finally, the diagonal entries of $B$ and $C$ are chosen such that all columns sum to the same value (this fixes all but one diagonal element, which can be chosen freely).

\subsection{Class CI$_{+-}$, $R_+R_+^{-T}=S^2=+1$, $R_-R_-^{-T}=-1$}
\label{subsec:class_CIpm}

Class CI$_{+-}$ is labeled by the symmetries $R_+R_+^{-T}=S^2=+1$ and $R_-R_-^{-T}=-1$. We can take the parametrization of AI$_+$ matrix, which has the $S^2=+1$ symmetry and, on top, require the $R_-R_-^{-T}=-1$ symmetry.
We implement the $R_-$ symmetry as $R_-=Y\otimes \id$. Equation~(\ref{eq:symm_C-}) reads as
\begin{equation}
	R_- M^T R_-^{-1}
	=
	\begin{pmatrix}
		0 & \id \\ -\id & 0
	\end{pmatrix}
	\begin{pmatrix}
		0 & B^T \\ A^T & 0
	\end{pmatrix}
	\begin{pmatrix}
		0 & -\id \\ \id & 0
	\end{pmatrix}
	=
	\begin{pmatrix}
		0 & -A^T\\ -B^T & 0
	\end{pmatrix}
	=
	\begin{pmatrix}
		0 & -A\\ -B & 0
	\end{pmatrix}
	=-M,
\end{equation}
which implies that $A=A^T$ and $B=B^T$.
Finally, the existence of the $S$ and $R_-$ symmetries automatically implies the existence of the third one, $R_+$, which is given by $R_+=X\otimes \id$. Indeed, the matrix $M$ we found above satisfies Eq.~(\ref{eq:symm_C+}):
\begin{equation}
	R_+ M^T R_+^{-1}
	=
	\begin{pmatrix}
		0 & \id \\ \id & 0
	\end{pmatrix}
	\begin{pmatrix}
		0 & B^T \\ A^T & 0
	\end{pmatrix}
	\begin{pmatrix}
		0 & \id \\ \id & 0
	\end{pmatrix}
	\\=
	\begin{pmatrix}
		0 & A^T\\ B^T &  0
	\end{pmatrix}
	=
	\begin{pmatrix}
		0 & A\\ B & 0
	\end{pmatrix}
	=M.
\end{equation}
Note that this corresponds to Eq.~(\ref{eq:parametrization_BDIdg}) with $A=0$ (because we need vanishing diagonal blocks to implement a $S^2=1$ symmetry). Hence, to apply Eq.~(\ref{eq:symm_cond_T-}), we can follow the same procedure as described there.
The parameterization of a Markov generator in class CI$_{+-}$ is thus:
\begin{equation}
	\label{eq:parametrization_CI+-}
	M=\begin{pmatrix}
		0 & A \\ B & 0
	\end{pmatrix},
	\qquad
	\begin{cases}
		A=A^T
		\\
		B=B^T
	\end{cases},
\end{equation}
where the diagonal elements of $A$ and $B$ are fixed as described after Eq.~(\ref{eq:parametrization_BDIdg}).

\subsection{Class BDI$_{++}$, $R_+R_+^{-T}=R_-R_-^{-T}=S^2=+1$}
\label{subsec:class_BDIpp}

The final example is that of class BDI$_{++}$, labeled by the symmetries $R_+R_+^{-T}=R_-R_-^{-T}=S^2=+1$, which requires a $4\times4$ block structure. We take $S=Z\otimes Z\otimes \id$, $R_+=X\otimes X\otimes \id$, and $R_-=Y\otimes Y\otimes \id$. Following the same procedure as before, we arrive at
\begin{equation}
	M=\begin{pmatrix}
		0 & A    & C   & 0   \\
		B & 0    & 0   & C^T \\
		D & 0    & 0   & A^T \\
		0 & D^T  & B^T & 0   \\
	\end{pmatrix}.
\end{equation}
Finally, to apply Eq.~(\ref{eq:symm_cond_T-}), we note that it can be written as a linear system of $4N$ equations (of which only $4N-1$ are independent) with $4N^2$ entries of $A$, $B$, $C$, and $D$ as variables. We take the dependent variables to be $B_{NN}$, $C_{iN}$, $C_{Ni}$, $D_{iN}$, and $D_{Ni}$, $i=1,\dots,N$. Choosing a large enough positive value for the sum of the columns, we can fix all other entries freely, and then solve for the dependent variables.

\subsection{Absence of analytic examples for the other classes}
For the remaining classes, the positivity constraint sets too many entries of $M$ to zero, which induces additional symmetries: either (i) originally-absent $S$ and $R_\pm$ symmetries, or (ii) commuting symmetries that break the desired $S$ or $R_\pm$ symmetry; in either case, the symmetry class is changed. 

As an illustrative example of case (i), we consider class CI, characterized by $R_-R_-^{-T}=-1$, $R_+R_+^{-T}=S^2=0$. If we implement $R_-=Y\otimes\id$, then, proceeding as before, we find the matrix representation
\begin{equation}
\label{eq:parametrization_CI}
	M=\begin{pmatrix}
          A & B \\ C & -A^T 
          \end{pmatrix},
	\qquad
	\begin{cases}
		B=B^T
		\\
		C=C^T
	\end{cases}.
\end{equation}
Because of the positivity of the entries, we must set $A=0$. However, this reduces Eq.~(\ref{eq:parametrization_CI}) to Eq.~(\ref{eq:parametrization_CI+-}), i.e., induces additional symmetries $R_+=X\otimes \id$ and $S=Z\otimes \id$, and gives, instead a parametrization of class CI$_{+-}$.

As an example of case (ii), we turn to class AI$_-$, characterized by $S^2=-1$ and $R_+R_+^{-T}=R_-R_-^{-T}=0$. Choosing $S=Y\otimes\id$, we obtain
\begin{equation}
\label{eq:parametrization_AIm}
	M=\begin{pmatrix}
           0 & A \\ A & 0 
          \end{pmatrix}.
\end{equation}
While this matrix has no additional $R_\pm$ symmetries (which would change its class), it admits a second $S$ symmetry, given by $S'=Z\otimes\id$ (as is evident from the chiral structure of $M$). Now, as discussed in Sec.~\ref{sec:classification}, two symmetries of the same type lead to a matrix that commutes with $M$,  this case $U=X\otimes\id$ (which, up to a phase, gives the product $SS'$, and, consequently, $M$ block diagonalizes in the basis where $U$ is diagonal). Crucially, $U$ \emph{anticommutes} with both $S$ and $S'$ and, hence, $S$ and $S'$ connect states with opposite eigenvalue of $U$ and do not act inside a single block of $M$. As a consequence, the irreducible blocks of $L$ do not have any symmetry and belong to class AI instead of AI$_-$.

The discussion of the preceding two examples can be systematized. We have checked that for operators $R_\pm$ and $S$ implemented by any string of up to four Pauli matrices (corresponding to matrices with up to a $16\times16$ block structure), no classes beyond the five discussed in Secs.~\ref{subsec:class_AI}--\ref{subsec:class_BDIpp} can be built in the computational basis.

\section{Systematic numerical solutions of the symmetry classes}
\label{sec:systematic}

We have seen that, working in the computational basis, only five of the fifteen classes of Markov matrices can be realized. In the following, we look for the existence of a different basis where $L$ has real and positive entries. To this end, we develop a simple but systematic stochastic optimization algorithm.

\subsection{Parametrization of the symmetry operators}

Let us start by choosing a convenient general parametrization of the symmetry operators $S$ and $R_\pm$. 
We first consider the classes with $S$ symmetry only. Without loss of generality, because $S^2=\pm\id$, we can write $S=W \Sigma_S W^{-1}$, where $W$ is an arbitrary invertible matrix with fixed norm, $\Tr W^T W=1$, and
\begin{equation}
	\begin{cases}
	\Sigma_S=\mathrm{diag}(\underbrace{+1,\dots,+1}_n,\underbrace{-1,\dots,-1}_{N-n}), \quad &\text{if $\eta_S=+1$},
	\\
	\Sigma_S=\underbrace{Y\oplus\cdots\oplus Y}_{N/2}, \quad &\text{if $\eta_S=-1$},
	\end{cases}
\end{equation}
with $\mathrm{diag}$ denoting a diagonal matrix.

If, instead, there is only a $R_\pm$ symmetry, implemented by either a real symmetric ($\eta_\pm=+1$) or a real antisymmetric ($\eta_\pm=-1$) matrix, without loss of generality, we can write it as $R_\pm= W \Sigma_\pm W^T$, with $W$ now orthogonal and
\begin{equation}\label{eq:param_Sigma}
	\begin{cases}
		\Sigma_\pm=\mathrm{diag}(s_1,\dots,s_N)\quad &\text{if $\eta_\pm=+1$},
		\\
		\Sigma_\pm=s_1 Y \oplus \dots \oplus s_{N/2}Y\quad &\text{if $\eta_\pm=-1$},
	\end{cases}
\end{equation} 
where $s_j$ are real numbers. Because of the homogeneity of the symmetry conditions, we can fix $\Tr R_\pm R_\pm^T=\sum_j s_j^2=1$.

Finally, we turn to the case with all three symmetries. We have seen that, in that case, we do not need to consider $S$ explicitly, as its square is contained in the compatibility condition~(\ref{eq:comm_RpRm}). In this case, we parametrize $\Sigma_+$ as in Eq.~(\ref{eq:param_Sigma}) and define $\Sigma_-$ through $R_-=W \Sigma_- W^T$. Note that, because we are using the \emph{same} $W$, $\Sigma_-$ does not need to satisfy Eq.~(\ref{eq:param_Sigma}), but it still satisfies $\Sigma_-^T=\eta_-\Sigma_-$. Plugging the definitions of $\Sigma_\pm$ into Eq.~(\ref{eq:comm_RpRm}) yields
\begin{equation}\label{eq:comm_Sigma}
	\Sigma_+ \Sigma_-^{-T}=\epsilon \Sigma_- \Sigma_+^{-T}.
\end{equation}
For example, for class BDI$_{-+}$ ($R_+R_+^{-T}=R_-R_-^{-T}=+1$, $S^2=-1$), Eq.~(\ref{eq:comm_Sigma}) requires that only $N/2$ $s_j$ are independent, and that $\Sigma_\pm$ take the form $\Sigma_+=s_1 Z\oplus\cdots\oplus s_{N/2} Z$ and $\Sigma_-=g\widetilde{\Sigma}_-$ with $\widetilde{\Sigma}_-=s_1 X\oplus \cdots \oplus s_{N/2} X$. Here, $g=(\pm I)\oplus\cdots \oplus(\pm I)$, where $I$ the two-dimensional identity and all signs can be independently chosen. By reorganizing basis elements, $g$ can be brought to the form $g=\mathrm{diag}(\underbrace{+1,\dots,+1}_n,\underbrace{-1,\cdots,-1}_{N/2-n})\otimes I$, so that we only need to specify the number $n$ of $+1$ entries. We have also assumed that all independent $s_j$ are nondegenerate. We can proceed similarly to find the general parametrization of the symmetry operators in the remaining classes. The result for the eight classes with all three symmetries is listed in Table~\ref{tab:Sigma}.

\begin{table}[t]
	\caption{\justifying Parametrization of the symmetry operators for the eight classes with all three symmetries.}
	\begin{tabular}{@{}lcccll@{}}
		\toprule
		Class      & $\eta_+$ & $\eta_-$ & $\epsilon$ & $\Sigma_+$                                                    & $\widetilde{\Sigma}_-$                                         \\ \midrule
		BDI$_{++}$ & $+1$     & $+1$     & $+1$       & $\mathrm{diag}(s_1,\dots,s_N)$                                & $\mathrm{diag}(s_1,\dots,s_N)$                                \\
		CI$_{--}$  & $+1$     & $-1$     & $+1$       & $s_1 I \oplus \cdots \oplus s_{N/2} I$                        & $s_1 Y \oplus \cdots \oplus s_{N/2} Y$                       \\
		BDI$_{--}$ & $-1$     & $+1$     & $+1$       & $s_1 Y \oplus \cdots \oplus s_{N/2} Y$                        & $s_1 I \oplus \cdots \oplus s_{N/2} I$                        \\
		CI$_{++}$  & $-1$     & $-1$     & $+1$       & $s_1 Y \oplus \cdots \oplus s_{N/2} Y$                        & $s_1 Y \oplus \cdots \oplus s_{N/2} Y$                        \\
		BDI$_{-+}$ & $+1$     & $+1$     & $-1$       & $s_1 Z \oplus \cdots \oplus s_{N/2} Z$                        & $s_1 X \oplus \cdots \oplus s_{N/2} X$                        \\
		CI$_{+-}$  & $+1$     & $-1$     & $-1$       & $s_1 Z \oplus \cdots \oplus s_{N/2} Z$                        & $s_1 Y \oplus \cdots \oplus s_{N/2} Y$                        \\
		BDI$_{+-}$ & $-1$     & $+1$     & $-1$       & $s_1 Y \oplus \cdots \oplus s_{N/2} Y$                        & $s_1 Z \oplus \cdots \oplus s_{N/2} Z$                        \\
		CI$_{-+}$  & $-1$     & $-1$     & $-1$       & $s_1 (Y\otimes I) \oplus \cdots \oplus s_{N/4} (Y \otimes I)$ & $s_1 (Z\otimes Y) \oplus \cdots \oplus s_{N/4} (Z \otimes Y)$ \\ \bottomrule
	\end{tabular}
	\label{tab:Sigma}
\end{table}

\subsection{Solution of the symmetry constraints}

For fixed symmetry operators $S$ and $R_\pm$, the symmetry transformations (\ref{eq:def_S})--(\ref{eq:def_Rm}) together with the Markov condition~(\ref{eq:def_Markov_3}) form a linear system for the entries of $L$ that we can solve exactly. 
To this end, we (row-) vectorize 
\begin{equation}
	L\mapsto |L\rangle=(L_{11},L_{12},\dots,L_{1N},L_{21},\dots,L_{NN})^T,
\end{equation}
in which case, Eqs.~(\ref{eq:def_S})--(\ref{eq:def_Rm}) read as, respectively,
\begin{align}
&\(S\otimes S^{-T}+\id_{N^2}-\frac 2N |\id_N\rangle\langle \id_N|\)|L\rangle=0,
\label{eq:vect_S}
\\
&\((R_+\otimes R_+^{-T})\mathcal{S}-\id_{N^2}\)|L\rangle=0,
\label{eq:vect_Rp}
\\
&\((R_-\otimes R_-^{-T})\mathcal{S}+\id_{N^2}-\frac 2N |\id_N\rangle\langle \id_N|\)|L\rangle=0,
\label{eq:vect_Rm}
\end{align}
where $\mathcal{S}$ is the \textsc{swap} operator, $\mathcal{S}(A\otimes B)\mathcal{S}=B\otimes A$ for arbitrary operators $A$ and $B$, which implements transposition of $L$. On the other hand, Eq.~(\ref{eq:def_Markov_3}) becomes
\begin{equation}
	\(\mathbf{e}_N^T\otimes \id_N\)|L\rangle=0,
\end{equation}
where $\mathbf{e}_N$ is the $N$-dimensional (column) vector with all entries equal to one. We can write the combined system of equations as $\mathcal{F}|L\rangle=0$. This is an overdetermined homogeneous system of equations with $N^2$ variables and $N^2+N$ (for classes with one symmetry) or $N^2+2N$ (for classes with three symmetries) equations (recall that, in the case of three symmetries, Eq.~(\ref{eq:vect_S}) is already implied by Eqs.~(\ref{eq:vect_Rp}) and (\ref{eq:vect_Rm}) and can be omitted). Nevertheless, for large enough $N$ ($N\geq2$, $4$, or $6$ depending on the class), there is always a nonzero family of solutions. 
However, a generic member of this family does not satisfy the positivity condition~(\ref{eq:def_Markov_2}). It is not even guaranteed that, for a given $\mathcal{F}$ (i.e., given $S$ and $R_\pm$), there exists any member satisfying it.

To proceed, we construct a basis $\{|b_j\rangle\}_{j=1}^K$ for the $K$-dimensional kernel of $\mathcal{F}$. We write a solution of the homogeneous system as $|L\rangle=\sum_{j=1}^K \ell_j |b_j\rangle$, where, without loss of generality, we can fix $\sum_j \ell_j^2=1$. We introduce a cost function
\begin{equation}
	f:=\frac{1}{N}\sum_{j\neq k}^N \frac{|L_{jk}|-L_{jk}}{2},
\end{equation}
which satisfies $f=0$ for a Markov matrix and $f>0$ for a non-Markov matrix, with a larger value signaling a larger departure from Markovianity, and look for the $\bm{\ell}$ that minimizes it:
\begin{equation}
	\bm{\ell}_*=\underset{\bm{\ell}\in S^K}{\mathrm{argmin}}\ f(\bm{\ell}).
\end{equation}

\subsection{Monte Carlo sampling}

We now use a stochastic optimization algorithm to minimize the cost function $f$. For a fixed class, we start from a random initial set $\{W^{(0)},\bm{s}^{(0)}\}$ and a choice of $g$. 
We use a (zero-temperature) Metropolis algorithm to decrease $f$, where at each step we perturb $\{W,\bm{s}\}$ with strength $\delta$. More precisely, we perturb
\begin{equation}
	\bm{s}^{(i)}\to \bm{s}^{(i+1)}=\frac{\bm{s}^{(i)}+\delta\ \bm{v}}{|\bm{s}^{(i)}+\delta\ \bm{v}|},
\end{equation}
where $\bm{v}$ is a vector of Gaussian random variables with zero mean and unit variance. $W$ is perturbed differently depending on whether it is orthogonal (classes with an $R_\pm$ symmetry) or not (classes with only $S$ symmetry). In the former, we use
\begin{equation}
	W^{(i)}\to W^{(i+1)}=e^{\delta A} W^{(i)},
\end{equation}
where $A$ is a Gaussian real antisymmetric random matrix, whereas in  the latter, we set
\begin{equation}
	W^{(i)}\to W^{(i+1)}=\frac{W^{(i)}+\delta\ V}{||W^{(i)}+\delta\ V||},
\end{equation}
where $V$ is a real Gaussian random matrix and $||\cdot||$ is the Hilbert-Schmidt norm. For each $\{W^{(i)},\bm{s}^{(i)}\}$, we compute the cost function $f$ and, if it is lower than that of the previous iteration, we accept the new configuration. After $m_\delta$ iterations without an acceptance, we decrease $\delta\to\delta/2$. 

\subsection{Results}

\begin{figure}[t]
    \centering
    \includegraphics[width=\textwidth]{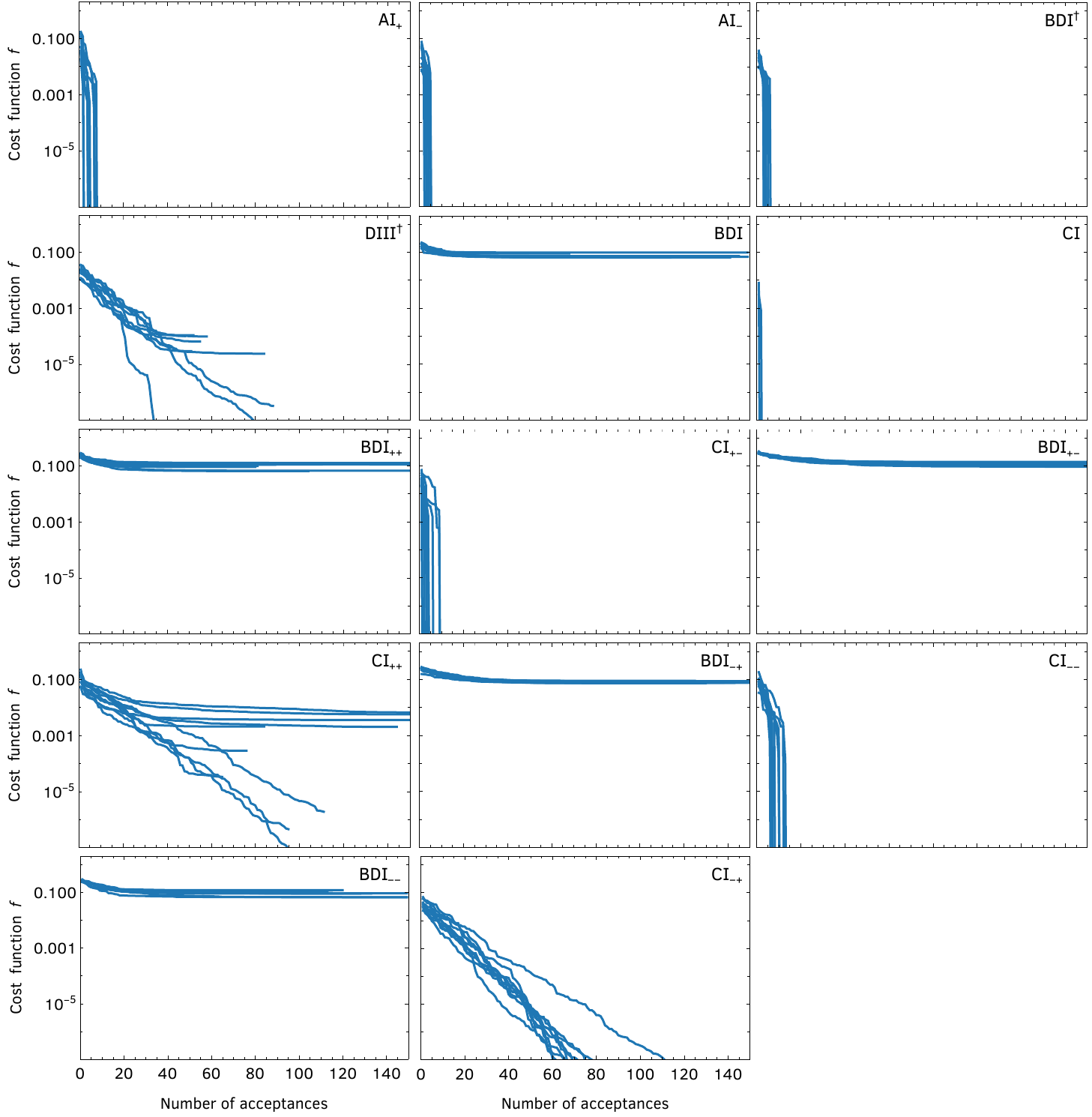}
    \caption{Results of the stochastic optimization algorithm for all fourteen classes with symmetry constraints, for $N=8$ and $n=2$. Each panel shows the value of the cost function $f$ as a function of the number of acceptances starting from 10 different initial conditions $\{W^{(i)},\bm{s}^{(i)}\}$, with $\delta=1$, $m_\delta=500$, and up to $2\times10^4$ iterations.}
    \label{fig:MC_results}
\end{figure}

Figure~\ref{fig:MC_results} depicts the results of the stochastic optimization algorithm for $N=8$, $n=2$ for all classes except AI (which has no symmetry constraints and therefore no algorithm to run). We started the algorithm with $\delta=1$, ran it for $2\times 10^4$ steps (or until $f<10^{-12}$ was reached), and chose $m_\delta=500$.
In nine of the classes (plus AI), there are solutions that satisfy the positivity condition. Of these, for six---AI$_+$, AI$_-$, BDI$^\dagger$, CI, CI$_{+-}$, and CI$_{--}$---$f=0$ is reached exactly after a finite (and small) number of iterations of the algorithm. For the other three---DIII$^\dagger$, CI$_{++}$, and CI$_{-+}$---$f$ decreases arbitrarily with the number of iterations but we have not reached $f<10^{-12}$ for any of our runs. Nevertheless, the results for these three classes are strikingly different from the five classes---BDI, BDI$_{++}$, BDI$_{+-}$, BDI$_{-+}$, and BDI$_{--}$---where $f$ plateaus at a constant of order $1$ and does not decrease with the number of iterations. We interpret this as a solution that is reached asymptotically for the former three classes, while no solution is found for the latter five. Remarkably, for one of the classes where we could not find a numerical solution (namely, BDI$_{++}$), there is an analytical solution given in Sec.~\ref{subsec:class_BDIpp}. This shows that not finding a numerical solution is a limitation of the algorithm and does not strictly rule out the existence of the four remaining classes. We checked extensively that different choices of sampling of $\bm{s}$ and $W$, values of $n$, acceptance criteria (including finite-temperature algorithms with different annealing schedules and varying initial values of $\delta$ and $m_\delta$), cost functions, and allowing for complex entries lead to qualitatively similar results; in particular, the classes for which solutions exist or not and whether $f=0$ is reached asymptotically or in a finite number of steps did not depend on any of these choices. 

We conclude with some empirical observations extracted from the results in Fig.~\ref{fig:MC_results} and Table~\ref{tab:classes}: no solution was found in any class with $R_-R_-^{-T}=+1$; in all classes with $R_-R_-^{-T}\neq+1$ and $R_+R_+^{-T}=-1$, the solution was reached asymptotically; in all other classes (i.e., those with $R_-R_-^{-T}\neq+1$ and $R_+R_+^{-T}\neq-1$), a solution was found exactly after a finite number of iterations. While we do not have a theoretical understanding of these features of the algorithm, they could help in devising new ones that could, potentially, show the existence of examples in the four missing classes.

\section{Examples and physical consequences}
\label{sec:physical_consequences}

\subsection{Example: Bipartite graph}
\label{sec:example_bipartite}

We now discuss simple physical examples of the families built in Sec.~\ref{sec:families}. The simplest example of a Markov generator in a nontrivial class is that of an exclusion process on a bipartite graph. Let us consider a graph whose $2N$ nodes are either black ($\bullet$) or white ($\circ$) and that there is no hopping between two nodes of the same type. Let us also denote the $j$th node of type $\bullet$ ($\circ$) as $\bullet_j$ ($\circ_j$). 
In the absence of any further constraints, $S=Z\otimes\id$ plays the role of sublattice symmetry and the Markov matrix of the stochastic process is given by
\begin{equation}
M=\begin{pmatrix}
0 & A \\ B & 0 
\end{pmatrix},
\end{equation}
where $A$ and $B$ are $N\times N$ real matrices containing the transition rates $\bullet\to\circ$ and $\circ\to\bullet$, respectively (the transitions $\bullet\to\circ$ and $\circ\to\bullet$ are not necessarily symmetric). 
The sum of the $k$th column of $M$ gives the rate at which particles hop into the $k$th node of the graph, coming from any other node. Imposing Eq.~(\ref{eq:symm_cond_T-}) thus amounts to fixing this total incoming rate into a node to be the same for all nodes. The generator $L$ then has $S^2=+1$ symmetry.
In the absence of any further restrictions, $L$ belongs to class AI$_+$.

If, we further impose that the hopping rate $\bullet_j\to\circ_k$ is the same as $\bullet_k\to\circ_j$, and similarly, the rates $\circ_j\to\bullet_k$ and $\circ_k\to\bullet_j$ coincide, then $A=A^T$ and $B=B^T$. (Note that we are \emph{not} assuming any symmetry between the hoppings $\bullet_j\to\circ_k$ and $\circ_j\to\bullet_k$, i.e., $A\neq B^T$.) Thus, in addition to the $S$ symmetry, we also have $R_+=X \otimes \id$, satisfying $R_+R_+^{-T}=+1$. The simultaneous existence of $S$ and $R_+$ symmetries automatically implies the existence of a $R_-$ symmetry, implemented here by $R_-=Y\otimes\id$, $R_-R_-^{-T}=-1$. Keeping the total incoming rate into any node still fixed, we have a Markov generator of class CI$_{+-}$.

If we allow for hopping between nodes of the same type, but which are restricted by the special property that the rate $\bullet_j\to \bullet_k$ is equal to the (reverse) rate $\circ_k\to \circ_j$, then we can write the Markov matrix $M$ as
\begin{equation}
M=\begin{pmatrix}
C & A \\ B & C^T
\end{pmatrix},
\qquad
\begin{cases}
A=A^T\\B=B^T
\end{cases},
\end{equation}
where $C$ contains the $\bullet\to\bullet$ and $\circ\to\circ$ rates. The diagonal blocks break the $S$ and $R_-$ symmetries, but preserve the $R_+$ symmetry, provided that the total incoming rate into every node is still the same. In that case, the Markov generator $L$ belongs to class BDI$^\dagger$.

Finally, we would need to consider a quadripartite graph to get the other class with an explicit parametrization, BDI$_{++}$.

\subsection{Dihedral symmetry of the spectrum}

Having presented some examples of stochastic processes in different classes in the previous section, we now turn to the spectral consequences of different symmetries. Because $L$ is always real, its eigenvalues are real or come in complex conjugate pairs. Likewise, the presence of an $S$ or $R_\pm$ symmetry also pairs different eigenvalues, as we now show. 

Let $\lambda_j$, $j=1,\dots,N$, be the eigenvalues of $L'$ and $\phi_j$ and $\phit_j$ the corresponding right (column) and left (row) eigenvectors, i.e.,
\begin{align}
\label{eq:evprob}
    &L'\phi_j=\lambda_j\phi_j,
    \\
    &\phit_j L'=\lambda_j \phit_j
\end{align}
Since $L'$ is real,  all eigenvalues are either real or come in pairs $(\lambda_j,\lambda_j^*)$.
Likewise, multiplying Eq.~(\ref{eq:evprob}) by $S$ from the left and using Eq.~(\ref{eq:def_S}), we have
\begin{equation}
    L'S\phi_j=-\lambda_jS\phi_j,
\end{equation}
that is, $S$ reflects the spectrum across the origin since $S\phi_j$ is an eigenvector of $L'$ with the eigenvalue $-\lambda_j$. Consequently, all eigenvalues are zero or come in pairs $(\lambda_j,-\lambda_j)$. Combined with the reflection of the spectrum across the real axis due to the reality of $L'$, it implies dihedral symmetry of the spectrum of $L'$, analogous to that observed in the quantum case~\cite{prosen2012PRL,prosen2012PRA,sa2023PRX}.

On the other hand, multiplying Eq.~(\ref{eq:evprob}) by $R_\pm$ from the left and using Eqs.~(\ref{eq:def_Rp}) or (\ref{eq:def_Rm}), we have
\begin{equation}
    (R_\pm \phi_j)^TL' =\pm\lambda_j(R_\pm\phi_j)^T,    
\end{equation}
i.e, $(R_\pm\phi_j)^T$ is a \emph{left} eigenvector corresponding to the eigenvalue $\pm\lambda_j$. For $R_-$, eigenvalues thus always come in pairs $(\lambda_j,-\lambda_j)$, and as in the case of an $S$ symmetry, the spectrum displays dihedral symmetry. For $R_+$ no dihedral symmetry is present, as it relates eigenvectors with the same eigenvalues.

\begin{figure}
    \centering
    \includegraphics[width=\textwidth]{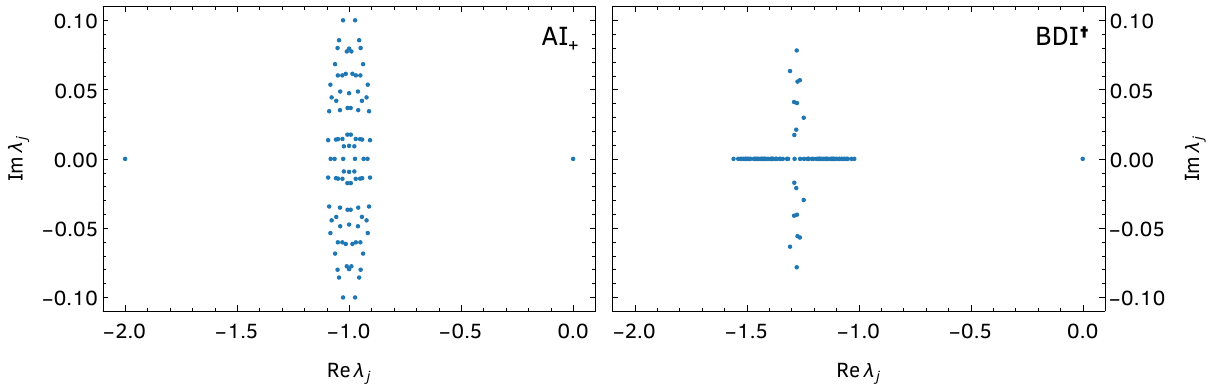}
    \caption{Spectrum of $L$ in the complex plane for a random Markov matrix from class AI$_+$ (left), sampled according to Eq.~(\ref{eq:parametrization_AI+}) and which displays dihedral symmetry, and class BDI$^\dagger$ (right), sampled from Eq.~(\ref{eq:parametrization_BDIdg}) and which does not show dihedral symmetry. In both cases, we set $N=100$ and the entries of the blocks $A$, $B$, $C$ are the absolute values of normal random variables with zero mean and variance $2/N$.}
    \label{fig:dihedral}
\end{figure}

To illustrate these properties, we show in Fig.~\ref{fig:dihedral} the spectra of a representative member of two classes with an analytical parametrization (see Sec.~\ref{sec:families}), namely, one with dihedral symmetry (class AI$_+$) and one without (class BDI$^\dagger$). 

\subsection{Kramers degeneracy}

Combining the results of the previous section about the reality of $L'$ and its transformation under $R_+$, it follows that for a given right eigenvector $\phi_j$, $(R_+\phi_j)^T$ is a \emph{left} eigenvector corresponding to the \emph{same} eigenvalue $\lambda_j$. But $\phit_j$ is also a left eigenvector associated with the eigenvalue $\lambda_j$. Now, by biorthogonality, we have $\phit_j \phi_j=1$. On the other hand, 
\begin{equation}
    (R_+ \phi_j)^T\phi_j 
    =  (R_+ \phi_j)^T (R_+^{-1}R_+\phi_j)
    =  (R_+^{-T}R_+ \phi_j)^T (R_+\phi_j)
    = \eta_+ \phi_j^T (R_+\phi_j),
\end{equation}
whence
\begin{equation}
(R_+\phi_j)^T\phi_j=0, 
\quad \text{if }\eta_+=-1.  
\end{equation}
Therefore, $\phit_j$ and $(R_+\phi_j)^T$ are linearly independent and every eigenvalue $\lambda_j$ is doubly degenerate, a result known as Kramers degeneracy in quantum theory and which we have shown to extend to the classical stochastic setting even when the antiunitarity conditions are relaxed. 

None of the examples for which we have analytical parametrization have $R_+R_+^T=-1$. Nevertheless, the numerical examples written out explicitly in Apps.~\ref{app:DIIIdg}, \ref{app:CIpp}, and \ref{app:CImp} show double degeneracy of all of their eigenvalues, as expected.

\subsection{Detailed balance and $R_+$ symmetry}

A Markov matrix $L$ satisfies detailed balance if there exist positive numbers $\pi_j$ such that 
\begin{equation} \label{eq:balance}
    L_{ij}\, \pi_j = L_{ji}\, \pi_i.
\end{equation}
Since this equation is linear in $\bm{\pi}$, we can assume without loss of generality that $\sum_j\pi_j=1$. If Eq.~\eqref{eq:balance} holds, $\bm{\pi}$ forms the stationary measure of the process defined by $L$ (because $\partial_t \pi_i=\sum_j L_{ij}\pi_j=\pi_i\sum_{j}L_{ji}=0$).
Clearly, a Markov matrix satisfying the detailed balance of Eq.~\eqref{eq:balance} possesses an $R_+$ symmetry as it is enough to choose
\begin{equation}
\label{eq:diagonal_Rp}
    R_+ := \mathrm{diag}(\pi_1,\pi_2,\dots,\pi_N)
\end{equation}
to ensure Eq.~\eqref{eq:def_Rp}, i.e. $R_+\, {L}^T= {L}\, R_+$ (recall that for $R_+$ the shift in $L$ is irrelevant so that Eq.~\eqref{eq:def_Rp} can be written in terms of $L$ instead of $L'$). 

More generally, we can define a vector $\bm{p}_*:=R_+ \mathbf{e}_N$ (where, as before, $\mathbf{e}_N$ is the $N$-dimensional vector with all entries equal to $1$), with component $(p_*)_i=\sum_j(R_+)_{ij}$. By construction, $\bm{p}_*$ is an invariant state of $L$ (because $\partial_t \bm{p}_*=LR_+\mathbf{e}_N=R_+L^T\mathbf{e}_N=0$). Assuming that $\bm{p}_*$ is the unique invariant state of $L$ (which should be the case for a generic Markov process), then, by the Perron-Frobenius theorem, all its components are positive, and $\bm{p}_*$ is an invariant measure. Thus, an $R_+$ symmetry can be thought of as nondiagonal detailed balance.

Finally, we note that detailed balance, Eq.~(\ref{eq:balance}) can be seen as a special case of self-duality of the Markov process described by $L$~\cite{schutz1994,schutz1997,giardina2009}. More generally, duality of Markov processes can be used, e.g., to write $m$-point correlation functions in $N$-particle systems in terms of correlation functions in $m$-particle systems~\cite{schutz1994,schutz1997}.

\subsection{Time reversal of stochastic trajectories}

Consider a Markov chain with states $i,j,\dots$ and the stochastic matrix $P=e^{\delta t\,L}$, for some time interval $\delta t$, whose entries are the transition probabilities $P_{ij}=\mathrm{Prob}(i\to j)$. The probability to follow a given stochastic trajectory $\bm{\omega}=(i_0,i_1,\dots,i_T)$ or its reverse $\bm{\omega}_\leftarrow=(i_T,i_{T-1},\dots,i_0)$ are given by, respectively,
\begin{equation}
\label{eq:trajectories}
    \dsP(\bm{\omega})=P_{i_0i_1}P_{i_1i_2}\cdots P_{i_{T-1}i_T}
    \qquad \text{and} \qquad
    \dsP(\bm{\omega}_\leftarrow)=P_{i_1i_0}P_{i_2i_1}\cdots P_{i_Ti_{T-1}}.
\end{equation}
In the presence of detailed balance, Eq.~(\ref{eq:balance}), which is inherited by $P$, we can immediately relate the trajectory $\bm{\omega}$ to its reverse through:
\begin{equation}
\label{eq:trajectories_diagonal}
    \dsP(\bm{\omega}_\leftarrow)=\frac{\pi_{i_T}}{\pi_{i_0}}\, \dsP(\bm{\omega}).
\end{equation}
This remark can simply be extended to the cases in which $R_+$ is an involutive permutation, up to a redefinition of the time-reversed trajectory.

In the preceding discussion, we have assumed the $R_+$ symmetry operator to be diagonal. A more general situation where a nondiagonal $R_+$ appears is when we do not have full knowledge of the trajectory, but only of some coarse-grained information $\mu$ at each time step (for instance, in the example of Sec.~\ref{sec:example_bipartite}, outcome $\mu$ could be whether a $\bullet\to\bullet$, $\circ\to\circ$, $\bullet\to\circ$, or $\circ\to\bullet$ transition occurred, without knowing exactly the initial and final states of the transition). Accordingly, we write
\begin{equation}
    P=\sum_\mu \Pi_\mu,
    \qquad (\Pi_\mu)_{ij}=\mathrm{Prob}(i\to j|\mu),
\end{equation}
where $P(i\to j|\mu)$ is the transition probability for a process $a\to b$ conditioned on the outcome $\mu$. The probability of a sequence of coarse-grained outcomes $\bm{\mu}=(\mu_1,\mu_2,\dots,\mu_T)$ is given by summing over all trajectories (\ref{eq:trajectories}) consistent with the outcomes $\bm{\mu}$:
\begin{equation}
    \mathbb{P}(i_0\to i_T|\bm{\mu})
    =\sum_{i_1,\dots,i_{T-1}} 
    (\Pi_{\mu_1})_{i_0 i_1} (\Pi_{\mu_2})_{i_1 i_2} \cdots (\Pi_{\mu_T})_{i_{T-1} i_T}
    =(\Pi_{\mu_1} \cdots \Pi_{\mu_T})_{i_0i_T}.
\end{equation}
(Here, we assume that the exact initial and final states $i_0$ and $i_T$ are known; if that is not the case, one performs an additional sum over them.)
The probability of the reverse sequence of outcomes $\bm{\mu}_\leftarrow=(\mu_T,\mu_{T-1},\dots,\mu_1)$ occurring is, accordingly,
\begin{equation}
     \mathbb{P}(i_T\to i_0|\bm{\mu}_\leftarrow)
    =\sum_{i_1,\dots,i_{T-1}} 
    (\Pi_{\mu_T})_{i_T i_{T-1}} \cdots (\Pi_{\mu_1})_{i_{1} i_0}
    =(\Pi_{\mu_1}^T\cdots \Pi_{\mu_T})_{i_0 i_T}.
\end{equation}
To proceed, we assume that all $\Pi_\mu$ have the same $R_+$ symmetry, $R_+ \Pi_\mu^T R_+^{-1}=\Pi_\mu$ (naturally, $R_+$ is also a symmetry of their sum, the stochastic matrix $P$). Then, we can write
\begin{equation}
\begin{split}
     \mathbb{P}(i_T\to i_0|\bm{\mu}_\leftarrow)
     &=(R_+^{-1}\Pi_{\mu_1}R_+ R_+^{-1} \Pi_{\mu_2}R_+ \cdots R_+^{-1}\Pi_{\mu_T}^T R_+)_{i_Ti_0}
     \\
     &=\sum_{j_0,j_T}(R_+^{-1})_{i_0j_0}(R_+)_{j_Ti_T}
     \mathbb{P}(j_0\to j_T|\bm{\mu}),
\end{split}
\end{equation}
i.e., the probability of a backward sequence of outcomes is given by the sum of the probabilities of all forward sequences compatible with $\bm{\mu}$ and weighted by the entries of the $R_+$ symmetry. In the case of a diagonal $R_+$, Eq.~(\ref{eq:diagonal_Rp}), we recover a conditioned version of Eq.~(\ref{eq:trajectories_diagonal}), namely
\begin{equation}
    \dsP(i_T\to i_0|\bm{\mu}_\leftarrow)=\frac{\pi_{i_T}}{\pi_{i_0}}\, \dsP(i_0\to i_T|\bm{\mu}).
\end{equation}

\subsection{Time reversal of correlation functions}

For the two symmetries that involve a minus sign, $R_-$ and $S$, the resulting operator $-L^T$ or $-L$ is not a proper generator of stochastic dynamics and, as such, the ``time-reversed'' trajectories are not properly defined. Nevertheless, as in the quantum case~\cite{sa2023PRX}, certain ``time-reversed'' correlation functions are still meaningful. For definiteness, we consider an $S$ symmetry and a vector of observables $\bm{O}$ and a probability vector $\bm{p}_0$ that are both invariant under it, $\bm{O}S=\bm{O}$ and $S\bm{p}_0=\bm{p}_0$. The time-dependent expectation value $\av{O(t)}=\bm{O}\cdot \bm{p}(t)$, where $\bm{p}(t)=e^{L t} \bm{p}_0$ satisfies:
\begin{equation}
    \av{O(t)}
    =\bm{O}\cdot e^{L t} \bm{p}_0
    =e^{-2\alpha t}\bm{O}S\cdot e^{-L t} S \bm{p}_0
    =e^{-2\alpha t}\av{O(-t)}.
\end{equation}
It follows that, even though the $S$-transformed matrix $SLS^{-1}$ does not generate Markov dynamics, the weighted expectation value $\av{\hat{O}(t)}=e^{\alpha t}\av{O(t)}$ is time-reversal symmetric, $\av{\hat{O}(t)}=\av{\hat{O}(-t)}$.

\section{Conclusions and outlook}

In this paper, we showed that a powerful classification scheme based on (anti) unitary symmetries in quantum mechanics can be extended to classical stochastic processes. By relaxing the assumption of antiunitarity, we put forward a fifteenfold classification based entirely on linear involutions of the Markov generator $L$ of the stochastic process. However, the existence of a preferred basis in which the off-diagonal matrix elements of $L$ are positive, strongly constrains the construction of analytical examples, a task that is usually straightforward in the quantum case. Because we were only able to parametrize $L$ analytically in five of the fifteen classes, we then turned to a numerical optimization algorithm to find examples in the remaining classes. To this end, we minimized the sum of negative off-diagonal elements under the constraint of the symmetry transformations, finding examples in six additional classes (either reaching a zero cost function in a finite number of steps or decreasing it asymptotically). Therefore, in total, we found representative examples in eleven of the fifteen classes. While we were not able to find any examples in the last four classes, neither do our approaches---both analytical or numerical---disprove their existence. As such, we leave proving no-go theorems for these classes or explicitly building examples in them as an open problem that deserves further investigation. We also defer the improvement of the Monte Carlo optimization algorithm such that system sizes $N\gg8$ can be reached efficiently to the future.

Our symmetry-based considerations allowed us to identify classical stochastic processes with so far unidentified properties, usually attributed solely to quantum processes, namely, the existence of dihedral spectral symmetry or Kramers' degeneracy. Furthermore, the former leads to a time-reversal-like property of expectation values of classical observables, even if the dynamics are out of equilibrium. Finally, we also interpreted $R_+$ as a generalized detailed balance and discussed its consequences for the stationary measure and stochastic trajectories of the associated Markov process. Our work opens the door for identifying these, and other unexplored, properties in realistic physical contexts.

\vspace{+2ex}

\noindent\textbf{Data availability.} Data generated during the current study is available from the corresponding author upon reasonable request.

\noindent\textbf{Competing interests.} The authors have no competing interests to declare that are relevant to the content of this article.

\noindent\textbf{Acknowledgments.} L.S.\ was supported by a Research Fellowship from the Royal Commission for the Exhibition of 1851. This work was supported by Fundação para a Ciência e a Tecnologia (FCT-Portugal) through Grant No. UID/CTM/04540/2019 (P.R.) and by the QuantERA II Project DQUANT funded through the European Union's Horizon 2020 research and innovation programme under Grant Agreement No. 101017733. T.P.\ acknowledges support from the European Research Council (ERC) through Advanced grant QUEST (Grant Agreement No. 101096208), and the Slovenian Research and Innovation Agency (ARIS) through the Program P1-0402. D.B.'s contribution was in part supported by the CNRS, the ENS, and the ANR project ESQuisses under contract number ANR-20-CE47-0014-01.

\begin{appendices}

\section{Steady-state blocks of reducible Markov matrices}
\label{app:Markov_steady_state}

In this appendix, we give a proof, with only a mild assumption, of the statement made in the main text that, for a reducible Markov matrix with a commuting symmetry, the block containing the steady state can also be reduced to Markov form, i.e., satisfies conditions (\ref{eq:def_Markov_1})--(\ref{eq:def_Markov_3}). We give evidence for the accuracy of the assumption behind the proof by working out in detail an example of a stochastic process with SU(2) symmetry, where it is naturally satisfied. 

We consider a reducible Markov matrix with states $\{|i\rangle\}_{i=1}^N$ and a commuting symmetry $LU=UL$. Let $|\bm{p}_\infty\rangle$ be the (generically unique) steady state, $\langle \bm{1}|=\sum_{i}\langle i |$ the associated left eigenvector (with $\langle \bm{1}|\bm{p}_\infty\rangle=1$), and $Q_u$ be the projector into the steady-state block where $U$ has eigenvalue $u$, $Q_u U Q_u=uQ_u$; we have $Q_u|\bm{p}_\infty\rangle=|\bm{p}_\infty\rangle$ and $\langle \bm{1}|Q_u=\langle \bm{1}|$.
Now, let $\{|u_a\rangle\}_{a=1}^d$ be a complete basis of the steady-state block, where $d$ is its dimension, i.e., $Q_u=\sum_a |u_a\rangle\langle u_a|$, and define a new basis
\begin{equation}
    |a\rangle = \frac{|u_a\rangle}{\langle\bm{1}|u_a\rangle}
    \qquad \text{and} \qquad
    \langle \at| = \langle\bm{1}|u_a\rangle\langle u_a|,
\end{equation}
which also satisfies $\sum_a\langle \at|=\langle\bm{1}|$ and $Q_u=\sum_a |a\rangle\langle \at|$.
We assume a basis $|u_a\rangle$ exists such that $\langle i|u_a\rangle$ are real and positive.
In this basis, the projection $L_u=Q_uLQ_u$ into the steady-state block is given by
\begin{equation}
\label{eq:proj_L}
    (L_u)_{ab}=\langle \at| L | b \rangle
    =\sum_{ij} \langle \at|i\rangle L_{ij} \langle j | b \rangle
    =\sum_{i,j\neq i}\left( p_{a|i} L_{ij} p_{j|b} - p_{a|i} L_{ji} p_{i|b} \right),
\end{equation}
where we used the fact that for the full Markov matrix in the original basis $L_{ii}=-\sum_{j\neq i}L_{ji}$ and defined the conditional probabilities $p_{a|i}:=\langle \at |i \rangle$ and $p_{i|a}:=\langle i | a \rangle$. Since $\langle \bm{1}|=\sum_i\langle i|$ and $\langle \bm{1}|i\rangle=1$, we have $\sum_a p_{a|i}=1$ and $\sum_ip_{i|a}=1$. From this, it follows that $p_{a|i}\in [0,1]$.

At this point, $(L_u)_{ab}$ is not guaranteed to be of Markov form.
To proceed, we assume that the steady-state block corresponds to the symmetric sector of $U$ and, furthermore, that 
\begin{equation}
\label{eq:markov_assumption}
    p_{a|i}\in\{0,1\}.
\end{equation}
To motivate assumption~(\ref{eq:markov_assumption}), we consider three coupled two-level systems with (original) basis 
\begin{equation}
    \{|i\rangle \}_{i=1}^8=\{
    |\uparrow\uparrow\uparrow\rangle,
    |\uparrow\uparrow\downarrow\rangle,
    \dots,
    |\downarrow\downarrow\downarrow\rangle
    \}.
\end{equation}
If there is a $\mathbb{Z}_2$ spin-flip symmetry, the symmetric sector has basis 
\begin{equation}
    \{|\at\rangle\}_{a=1}^4=\{
    |\uparrow\uparrow\uparrow\rangle+|\downarrow\downarrow\downarrow\rangle,
    |\downarrow\uparrow\uparrow\rangle+|\uparrow\downarrow\downarrow\rangle,
    |\uparrow\downarrow\uparrow\rangle+|\downarrow\uparrow\downarrow\rangle,
    |\uparrow\uparrow\downarrow\rangle+|\downarrow\downarrow\uparrow\rangle
    \};
\end{equation}
If, instead, the system has SU(2) symmetry, the symmetric sector of maximum spin is generated by 
\begin{equation}
    \{|\at\rangle\}_{a=1}^4=\{
     |\uparrow\uparrow\uparrow\rangle,
     |\uparrow\uparrow\downarrow\rangle+|\uparrow\downarrow\uparrow\rangle+|\downarrow\uparrow\uparrow\rangle,
    |\uparrow\downarrow\downarrow\rangle+|\downarrow\uparrow\downarrow\rangle+|\downarrow\downarrow\uparrow\rangle,
    |\downarrow\downarrow\downarrow\rangle
    \};
\end{equation}
both bases satisfy assumption~(\ref{eq:markov_assumption}). In the next section, we work out the SU(2) example in detail.

Equipped with assumption~(\ref{eq:markov_assumption}), we can show that $L$ projected into the steady-state block, Eq.~(\ref{eq:proj_L}), indeed satisfies the conditions (\ref{eq:def_Markov_1})--(\ref{eq:def_Markov_3}) of a Markov matrix. First, because $p_{a|i}$, $p_{i|a}$, and $L_{ij}$ are all real, it immediately follows that $(L_u)_{ab}$ is also real. Next, to check the positivity of the nondiagonal entries of $L_u$, we note that if $a\neq b$, the last term in the right-hand side of Eq.~(\ref{eq:proj_L}) vanishes by assumption. Since $p_{a|i}$, $p_{i|a}$ and $L_{ij}$ with $i\neq j$ are all nonnegative, it follows that
\begin{equation}
    (L_u)_{ab}=\sum_{i,j\neq i}p_{a|i} L_{ij} p_{j|b}\geq 0, \quad \text{if }a\neq b.
\end{equation}
Finally, using $\sum_a p_{a|i}=1$, we have
\begin{equation}
    \sum_a(L_u)_{ab}
    =\sum_{i,j\neq i}\left(L_{ij} p_{ j | b}- L_{ji} p_{i | b}\right)=0,
\end{equation}
thus showing that probability is conserved, which concludes the proof of Eqs.~(\ref{eq:def_Markov_1})--(\ref{eq:def_Markov_3}) for $L_u$.

\subsection{Example: SU(2) symmetry}

Consider the following Markov process for the evolution of the probability vector $\ket{\bm{p}}=\sum_{i_{1}\cdots i_{N}}p_{i_{1},\dots, i_{N}}\ket{i_{1},\dots, i_{N}}$ over the states $|i_{1}\dots i_{N}\rangle$, $i_n\in\{-1/2,1/2\}$:
\begin{equation}
\begin{split}
L&=  \gamma_{+}\sum_{i}\left[\sigma_{i}^{+}+\frac{1}{2}\left(\sigma_{i}^{z}-1\right)\right]+\gamma_{-}\sum_{i}\left[\sigma_{i}^{-}-\frac{1}{2}\left(\sigma_{i}^{z}+1\right)\right]\\
 &+ \gamma_{0}\sum_{ij}\left[\sigma_{i}^{+}\sigma_{j}^{-}-\frac{1}{4}\left(\sigma_{i}^{z}+\sigma_{j}^{z}+\sigma_{i}^{z}\sigma_{j}^{z}-\frac{N(N+2)}{N^{2}}\right)\right],
 \end{split}
\end{equation}
where $\sigma_j^\alpha$, $\alpha=x,y,z$ are the usual Pauli matrices supported on site $j$, which describes classical spin injection and removal at each site at rates $\gamma_+$ and $\gamma_-$, respectively, and spin exchange between sites $i$ and $j$ at rate $\gamma_0$.
Introducing collective spin variables $S^{\alpha}=\sum_{i}\sigma_{i}^{\alpha}/2$, $L$ can be written as
\begin{equation}
\begin{split}
L&=  \gamma_{+}\left(S^{+}+S^{z}-\frac{N}{2}\right)
+\gamma_{-}\left(S^{-}-S^{z}-\frac{N}{2}\right)\\
  &+ \gamma_{0}\left[S^{+}S^{-}-S^{z}-\left(S^{z}\right)^{2}+\frac{N}{2}\left(\frac{N}{2}+1\right)\right].
\end{split}
\end{equation}
$L$ commutes with $S^{2}=(S^x)^2+(S^y)^2+(S^z)^2$ and, thus, can be decomposed into su(2) irreducible sectors. The steady state belongs to the symmetric sector of maximal spin, where $S^{2}=s\left(s+1\right)$ with $s=N/2$, which is spanned by the $2s+1$ states ($m\in\{-2s,\dots,2s\}$)
\begin{equation}
\ket{s,m} \propto\sum_{\sigma\in S_{2s}}\ket{i_{\sigma(1)},\dots,i_{\sigma(N)}}_m,
\end{equation}
with $|\,i_{1},\dots,i_{N}\rangle_m=|\underset{s+m}{\underbrace{1/2,1/2,1/2,\dots,}}\ \underset{s-m}{\underbrace{-1/2,-1/2,-1/2,\dots}}\,\rangle$,
and where the sum has $\left(\begin{array}{c}
2s\\
s+m
\end{array}\right)$ terms, i.e.,
\begin{equation}
\langle \bm{1}|s,m\rangle  =\sqrt{\left(\begin{array}{c}
2s\\
s+m
\end{array}\right)}.
\end{equation}
Therefore, we can define
\begin{equation}
\bra{\widetilde{m}} =\sqrt{\left(\begin{array}{c}
2s\\
s+m
\end{array}\right)}\bra{s,m}
\qquad \text{and} \qquad 
\ket m  =\frac{1}{\sqrt{\left(\begin{array}{c}
2s\\
s+m
\end{array}\right)}}\ket{s,m}.
\end{equation}
Using this basis, we find that the Markov matrix reduced to the symmetric subspace,
$S^{2}=s\left(s+1\right)$, is given by 
\begin{equation}
\begin{split}
\left[L_{s\left(s+1\right)}\right]_{m,m'}
&= \bra{\widetilde{m}}L\ket{m'}\\
&= \gamma_{+}\left[\left(s-m\right)\delta_{m,m'+1}-\left(s-m\right)\delta_{m,m'}\right] \\
 &+\gamma_{-}\left[\left(s+m\right)\delta_{m,m'-1}-\left(s+m\right)\delta_{m,m'}\right],
 \end{split}
\end{equation}
which is clearly a Markov matrix. 

Let us illustrate these results for the specific case $N=3$. The $\langle \widetilde{m}|$ basis for the symmetric sector is 
\begin{equation}
\begin{split}
\bra{\widetilde{m}=3/2} & =\bra{1/2,1/2,1/2},\\
\bra{\widetilde{m}=1/2} & =\bra{-1/2,1/2,1/2}+\bra{1/2,-1/2,1/2}+\bra{1/2,1/2,-1/2},\\
\bra{\widetilde{m}=-1/2} & =\bra{1/2,-1/2,-1/2}+\bra{-1/2,1/2,-1/2}+\bra{-1/2,-1/2,1/2},\\
\bra{\widetilde{m}=-3/2} & =\bra{-1/2,-1/2,-1/2}.
\end{split}
\end{equation}
and $\ket m=\ket{\widetilde{m}}/\braket {\bm{1}}{\widetilde{m}}$. An example of
the conditional probabilities is
\begin{equation}
\begin{split}
p_{i=\left(\frac{1}{2},-\frac{1}{2},\frac{1}{2}\right)|m=\frac{1}{2}} & =\braket{\frac{1}{2},-\frac{1}{2},\frac{1}{2}}{\frac{1}{2}}=\frac{1}{3},\\
p_{m=\frac{1}{2}|i=\left(\frac{1}{2},-\frac{1}{2},\frac{1}{2}\right)} & =\braket{\widetilde{\frac{1}{2}}}{\frac{1}{2},-\frac{1}{2},\frac{1}{2}}=1.
\end{split}
\end{equation}
Finally, the irreducible Markov matrix restricted to the steady-state sector is, thus,
\begin{equation}
L_{s\left(s+1\right)}= \begin{pmatrix}
-3\gamma_{+} & \gamma_{-} & 0 & 0\\
3\gamma_{+} & -\gamma_{-}-2\gamma_{+} & 2\gamma_{-} & 0\\
0 & 2\gamma_{+} & -2\gamma_{-}-\gamma_{+} & 3\gamma_{-}\\
0 & 0 & \gamma_{+} & -3\gamma_{-}
\end{pmatrix}.
\end{equation}

\medskip

\section{Explicit examples of Markov matrices in different classes}

In this appendix, we write down one specific example of a Markov matrix belonging to each of the classes obtained numerically with the stochastic optimization algorithm in Sec.~\ref{sec:systematic} for which no analytical solution exists. Recall that for classes AI$_-$, CI, and CI$_{--}$, the cost function reaches $0$ in a finite number of steps and, consequently, the matrices displayed below for those classes satisfy the Markov property to machine precision; for classes DIII$^\dagger$, CI$_{++}$, and CI$_{-+}$, $f\to0$ only asymptotically and, therefore, there are some residual nondiagonal negative entries in the matrices obtained. (For conciseness, all entries have been rounded to four decimal places.)

\subsection{Class AI$_-$}
\label{app:AIm}

\begin{equation}
L=\left(
\begin{array}{rrrrrrrr}
 -0.2911 & 0.0004 & 0.0011 & 0.0061 & 0.1413 & 0.0038 & 0.0882 & 0.0748 \\
 0.0167 & -0.2849 & 0.0587 & 0.0577 & 0.0246 & 0.0973 & 0.1081 & 0.0902 \\
 0.0538 & 0.0058 & -0.2157 & 0.0749 & 0.0004 & 0.0457 & 0.0040 & 0.0359 \\
 0.0118 & 0.0471 & 0.0655 & -0.1896 & 0.0419 & 0.0479 & 0.0130 & 0.1512 \\
 0.0382 & 0.0011 & 0.0012 & 0.0025 & -0.2928 & 0.0058 & 0.0026 & 0.0023 \\
 0.0010 & 0.0264 & 0.0664 & 0.0001 & 0.0838 & -0.3096 & 0.0976 & 0.1434 \\
 0.0008 & 0.0716 & 0.0134 & 0.0030 & 0.0006 & 0.0716 & -0.3142 & 0.0039 \\
 0.1688 & 0.1324 & 0.0092 & 0.0453 & 0.0001 & 0.0374 & 0.0006 & -0.5016 \\
\end{array}
\right)
\end{equation}

\subsection{Class DIII$^\dagger$}
\label{app:DIIIdg}

\begin{equation}
L=\left(
\begin{array}{rrrrrrrr}
 -0.2878 & 0.0004 & 0.0379 & 0.0048 & -0.0000 & 0.0381 & 0.0583 & 0.0292 \\
 0.0000 & -0.3474 & 0.0000 & 0.0000 & -0.0000 & 0.0000 & 0.0057 & 0.0112 \\
 0.0320 & 0.0344 & -0.2938 & 0.0017 & -0.0005 & 0.0320 & 0.0161 & 0.0003 \\
 0.0000 & 0.0056 & 0.0000 & -0.3157 & 0.0140 & -0.0000 & 0.0068 & 0.0090 \\
 0.0000 & 0.1883 & 0.0000 & 0.2985 & -0.0133 & -0.0000 & 0.0395 & 0.1655 \\
 0.2557 & 0.1095 & 0.2557 & 0.0107 & -0.0002 & -0.0702 & 0.2172 & 0.1193 \\
 0.0000 & 0.0001 & 0.0002 & 0.0000 & -0.0000 & 0.0000 & -0.3528 & 0.0000 \\
 0.0000 & 0.0090 & 0.0000 & 0.0000 & -0.0000 & 0.0000 & 0.0093 & -0.3345 \\
\end{array}
\right)
\end{equation}

\subsection{Class CI}
\label{app:CI}

\begin{equation}
L=\left(
\begin{array}{rrrrrrrr}
 -0.2780 & 0.0001 & 0.0488 & 0.0008 & 0.0395 & 0.0005 & 0.1236 & 0.1520 \\
 0.0019 & -0.2660 & 0.1728 & 0.0006 & 0.0003 & 0.0003 & 0.0499 & 0.0177 \\
 0.0446 & 0.0360 & -0.4147 & 0.0010 & 0.0006 & 0.2001 & 0.0006 & 0.0001 \\
 0.0068 & 0.0229 & 0.0023 & -0.1649 & 0.0059 & 0.0016 & 0.0256 & 0.0851 \\
 0.0028 & 0.0001 & 0.0364 & 0.0009 & -0.1926 & 0.0000 & 0.0312 & 0.0056 \\
 0.0566 & 0.0866 & 0.1449 & 0.0364 & 0.0016 & -0.2665 & 0.0811 & 0.1645 \\
 0.1644 & 0.0636 & 0.0001 & 0.0006 & 0.1072 & 0.0031 & -0.3177 & 0.0009 \\
 0.0010 & 0.0569 & 0.0094 & 0.1246 & 0.0375 & 0.0608 & 0.0057 & -0.4259 \\
\end{array}
\right)
\end{equation}

\subsection{Class CI$_{++}$}
\label{app:CIpp}

\begin{equation}
    L=\left(
\begin{array}{cccccccc}
 -0.2546 & 0.0546 & 0.0118 & 0.0000 & 0.0712 & 0.0154 & 0.0283 & 0.0005 \\
 0.0755 & -0.2116 & 0.0612 & -0.0000 & 0.0345 & 0.0741 & 0.0676 & 0.1150 \\
 0.0477 & 0.0013 & -0.4242 & 0.0000 & 0.0197 & 0.0770 & 0.1564 & 0.0111 \\
 0.0000 & -0.0000 & -0.0000 & -0.0000 & -0.0000 & 0.0000 & -0.0000 & 0.0000 \\
 0.0395 & 0.0099 & 0.0388 & 0.0000 & -0.3768 & 0.0428 & 0.0428 & 0.1743 \\
 0.0733 & 0.0004 & 0.1068 & 0.0000 & 0.0659 & -0.2469 & 0.0054 & 0.0493 \\
 0.0160 & 0.1382 & 0.1633 & 0.0000 & 0.0208 & 0.0006 & -0.3474 & 0.0586 \\
 0.0026 & 0.0070 & 0.0423 & 0.0000 & 0.1648 & 0.0370 & 0.0469 & -0.4090 \\
\end{array}
\right)
\end{equation}

\subsection{Class CI$_{--}$}
\label{app:CImm}

\begin{equation}
L=\left(
\begin{array}{rrrrrrrr}
 -0.3648 & 0.0452 & 0.0588 & 0.2540 & 0.0665 & 0.0359 & 0.0630 & 0.0580 \\
 0.0293 & -0.2365 & 0.0469 & 0.0174 & 0.0391 & 0.0531 & 0.0405 & 0.0403 \\
 0.0681 & 0.0542 & -0.1846 & 0.0791 & 0.0137 & 0.0570 & 0.0654 & 0.0248 \\
 0.1648 & 0.0002 & 0.0012 & -0.4525 & 0.0618 & 0.0507 & 0.0037 & 0.0424 \\
 0.0795 & 0.0050 & 0.0337 & 0.0091 & -0.3407 & 0.0278 & 0.0198 & 0.0201 \\
 0.0137 & 0.0383 & 0.0253 & 0.0438 & 0.0470 & -0.2993 & 0.0455 & 0.0720 \\
 0.0041 & 0.0660 & 0.0185 & 0.0222 & 0.0547 & 0.0146 & -0.2710 & 0.0551 \\
 0.0053 & 0.0276 & 0.0002 & 0.0269 & 0.0579 & 0.0602 & 0.0331 & -0.3126 \\
\end{array}
\right)
\end{equation}

\subsection{Class CI$_{-+}$}
\label{app:CImp}

\begin{equation}
    L=\left(
\begin{array}{cccccccc}
 -0.0000 & 0.0000 & -0.0000 & -0.0000 & -0.0000 & 0.0000 & 0.0000 & -0.0000 \\
 -0.0000 & -0.3941 & 0.0497 & 0.0433 & 0.0516 & 0.1812 & 0.0268 & 0.0415 \\
 -0.0000 & 0.0544 & -0.4025 & 0.0222 & 0.0544 & 0.0581 & 0.1834 & 0.0300 \\
 0.0000 & 0.0348 & 0.0151 & -0.2500 & 0.0561 & 0.0423 & 0.0494 & 0.0523 \\
 0.0000 & 0.0659 & 0.0533 & 0.0322 & -0.2447 & 0.0302 & 0.0242 & 0.0390 \\
 0.0000 & 0.1789 & 0.0540 & 0.0535 & 0.0168 & -0.3941 & 0.0367 & 0.0542 \\
 0.0000 & 0.0296 & 0.1835 & 0.0579 & 0.0309 & 0.0381 & -0.3726 & 0.0327 \\
 -0.0000 & 0.0305 & 0.0469 & 0.0410 & 0.0350 & 0.0442 & 0.0521 & -0.2496 \\
\end{array}
\right)
\end{equation}

\end{appendices}
 
\bibliography{bib}
	
\end{document}